\documentclass{elsart}
\usepackage{epsfig}
\usepackage{amssymb}

\newcommand{\bm}[1]{\mbox{\boldmath$#1$}}
\def\0{\phantom{0}}
\begin{document}

\baselineskip24pt
\begin{frontmatter}




\title{Joule-Thomson inversion curves of mixtures by molecular simulation in comparison to advanced equations of state: natural gas as an example} 
\author{Jadran Vrabec\corauthref{cor1}}
\corauth[cor1]{Corresponding author, Tel.: +49-711/685-66107, Fax: +49-711/685-66140}
\ead{vrabec@itt.uni-stuttgart.de}
\ead[url]{www.itt.uni-stuttgart.de}
\author{Ashish Kumar~~~~~~}
\author{Hans Hasse}
\address{Institut f\"ur Technische Thermodynamik und Thermische Verfahrenstechnik,
Universit\"at Stuttgart, D-70550 Stuttgart, Germany}

\begin{abstract}
Molecular modelling and simulation as well as four equations of state (EOS) are applied to natural gas mixtures regarding Joule-Thomson (JT) inversion. 
JT inversion curves are determined by molecular simulation for six different natural gas mixtures consisting of methane, nitrogen, carbon dioxide and ethane. These components are also regarded as pure fluids, leading to a total of ten studied systems. The results are compared to four advanced mixture EOS: DDMIX, SUPERTRAPP, BACKONE and the recent GERG-2004 Wide-Range Reference EOS. It is found that molecular simulation is competitive with state-of-the-art EOS in predicting JT inversion curves. The molecular based approaches (simulation and BACKONE) are superior to DDMIX and SUERTRAPP.
\bigskip
\begin{flushleft}
\end{flushleft}
\end{abstract}
\end{frontmatter}
\section{Introduction}
Due to the eminent importance of natural gas, knowledge on its thermodynamic behaviour and appropriate property models are of great interest. A property often needed for applications of natural gases is the adiabatic (or isenthalpic) Joule-Thomson (JT) coefficient $\mu$, which is defined as the derivative of the temperature $T$ with respect to the pressure $p$ at constant enthalpy $h$ and constant composition $\bm{x}$ \begin{equation}
\mu~=~\left( \frac{\partial T}{\partial p} \right)_{h,\bm{x}},
\end{equation}
or, using basic thermodynamics relations,
\begin{equation}
\mu~=~-\frac{1}{c_p}\left( \frac{\partial h}{\partial p} \right)_{T,\bm{x}},
\end{equation}
where $c_p$ is the isobaric heat capacity. The JT inversion curve, connecting all state points with $\mu$=0, divides the pressure-temperature plane into two regions. In the lower region, $\mu$ is positive so that an adiabatic expansion leads to a decrease in temperature. In the upper region, $\mu$ is negative. It can be shown that the cooling effect is maximized if an expansion starts from the inversion pressure.

The experimental determination of a fluid's JT inversion coefficient demands precise measurements of volumetric and caloric properties, while the JT inversion curve extends over a broad range of temperature and pressure. Temperature and pressure reach up to a five-fold and twelve-fold of their critical values, respectively. Experimental JT inversion curve data are therefore scarce, sometimes unreliable \cite{hiza}, and mostly available only for pure fluids \cite{perry}. 
An example for mixture data is the work of Charnley et al. \cite{charnley} on e.g. carbon dioxide + nitrous oxide, carbon dioxide + ethylene. A good overview over experimental data is given in \cite{perry}.

For industrial applications there is a need for a proper representation and also for the prediction of JT data, as it is not feasible to measure it for all relevant and often very different blends. 
Some authors, e.g., Miller \cite{miller} or Gunn et al. \cite{gunn}, have proposed direct representations of the JT inversion curve, which are correlations in terms of reduced temperature and pressure. They provide rule-of-thumb data for simple fluids, but have little predictive power. 
Significantly more valuable are proper equations of state (EOS) that contain much more thermodynamic information and are valid for a broad range of state points. 
Thus, extensive efforts are made to use EOS for predictions of the JT inversion curve. 
Examples for the use of cubic EOS are modified versions of Peng-Robinson \cite{haghighi1,haghighi2}, Redlich-Kwong \cite{haghighi1,juris}, Soave-Redlich-Kwong \cite{haghighi2}, Patel-Teja \cite{haghighi1} or other cubic EOS \cite{haghighi3} with varying parameter functions. 
Both type of cubic EOS and type of parameter function strongly influence the JT inversion curve, particularly in the high temperature region. A given combination might yield good results for a specific fluid, but fails for others \cite{haghighi1,haghighi2,juris,haghighi3}. Hence, it can be concluded that cubic EOS are not generally reliable for JT inversion curve predictions.

Among mixtures, natural gases are the ones that were investigated most extensively both experimentally and theoretically so that very reliable thermodynamic data and models are available. 
Therefore, natural gases are excellent test systems to validate thermodynamic models for mixtures. 

Natural gas from the rig is a mixture of typically seventeen components (containing methane, nitrogen, carbon dioxide, ethane, propane) \cite{bergman}, but usually its main component is methane \cite{us_compo,california}. 
As a natural product, it has a great variability in composition and, depending on conditions in the formation process, considerable quantities of nitrogen (up to 60 mole \%), carbon dioxide (up to mole 50 \%) or ethane (up to mole 20 \%)
are encountered \cite{us_compo}. 

For a number of pure natural gas components, reference EOS have been developed based on a vast experimental data set considering different thermodynamic properties, e.g., methane \cite{wagmeth}, nitrogen \cite{wagn2}, carbon dioxide \cite{wagco2} and ethane \cite{ethane_eos}. 
Reference EOS have an empirical background, but they are parameterized extremely carefully, taking also available experimental JT coefficients into account. 
Hence, they are regarded in this work as the best available information. 

For mixtures, the National Institute of Standards and Technology (NIST) \cite{nist} provided two classical phenomenological EOS, i.e. DDMIX \cite{ddmix1,ddmix2} and SUPERTRAPP \cite{supertrapp}.
DDMIX is an implementation of the NIST extended corresponding states model for mixtures, whereas SUPERTRAPP is based on both a modified Peng-Robinson EOS and the NIST extended corresponding states model for mixtures.
Both were parameterized to experimental pure substance and mixture data.  
Particularly SUPERTRAPP is often used in the literature as a property model for designing cooling cycles with mixed coolants, e.g. \cite{maytal,huangx}. 

There are also physically based EOS that take the different molecular interactions, like dispersion or polarity, explicitly into account; an example is the BACKONE-EOS \cite{backone}. 
Such EOS can be parameterized for real substances with a very small experimental data set, e.g. a few vapour-liquid equilibrium data points, as they have a good predictive power. 
Furthermore, the GERG-2004 Wide-Range Reference EOS \cite{newwagner} has become available recently, which applies the concept of reference EOS to mixtures. 
Here too, a vast set of experimental data, both pure substance and mixture, was used for the development.   

Molecular modelling and simulation offers an interesting alternative approach for predicting thermodynamic properties. 
Instead of describing those macroscopic properties directly, the intermolecular interactions are described. 
Previous work from our group \cite{vrabecjt} has demonstrated the good predictive power of molecular models for JT inversion curves for different pure fluids, but also for the mixture air which has been modelled as a ternary system containing nitrogen, oxygen and argon. 
The objective of the present paper is to validate the predictive power of  available molecular models by comparing the results to the four advanced EOS mentioned above using different natural gas mixtures and their most important pure components regarding JT inversion.

\section{Molecular model and simulation method}
Most publications on JT inversion curves by molecular simulation are based on the spherical Lennard-Jones (LJ) model 
\cite{bessi,heyes,collj1,gaurav,compress1,compress2,thermal,new1}, which is appropriate only for simple molecules like methane and the noble gases. 
However, this simple potential model is well suited to further develop molecular simulation techniques for determining JT inversion as reliable simulation data is available for comparison. E.g., Colina et al. \cite{colina} have chosen two different routes, i.e. via compressibility \cite{compress1,compress2} or via thermal expansivity \cite{thermal}, to simulate the JT inversion curve. 
Work on more complex fluids is still scarce, examples are Escobedo et al. \cite{new1} (nitrogen), Chac\'{i}n et al. \cite{mulco2} (carbon dioxide), L\'{i}sal
et al. \cite{lisal} (R32), Krist\'{o}f et al. \cite{kris} (hydrogen sulphide), or a recent work of our group \cite{vrabecjt} dealing with fifteen different pure substances (including methane, carbon dioxide, R134a, R143a, R152a) and the mixture air.
Two publications, dealing with multi-component natural gas mixtures, should be mentioned: Escobedo et al. \cite{new1}
predicted the JT inversion curve for a seven-component system, but compared it to cubic EOS data only. A favourable 
comparison between simulation and experiment was presented by Lagache et al. \cite{new2} regarding the JT inversion pressure
of a 20-component mixture for one specified temperature.
  
The effective 2CLJQ pair potential was used as molecular model here to describe the intermolecular interactions in all cases. 
This can be done, as only the four most common components methane, nitrogen, carbon dioxide and ethane were considered, being well suited for this modelling approach. 
The 2CLJQ potential is composed out of two identical Lennard-Jones sites a distance $L$ apart (2CLJ) and a point quadrupole with momentum $Q$ placed in the geometric centre of the molecule oriented along the molecular axis 
\begin{equation}
u_{\rm 2CLJQ}(\bm{r}_{ij},\bm{\omega}_i,\bm{\omega}_j,L,Q) = u_{\rm 2CLJ}(\bm{r}_{ij},\bm{\omega}_i,\bm{\omega}_j,L)+u_{\rm Q}(\bm{r}_{ij},\bm{\omega}_i,\bm{\omega}_j,Q), \label{u2CLJQ}
\end{equation}
wherein $u_{\rm 2CLJ}$ is the contribution of the four Lennard-Jones interactions
\begin{equation}
u_{\rm 2CLJ}(\bm{r}_{ij},\bm{\omega}_i,\bm{\omega}_j,L)=\sum_{a=1}^{2} \sum_{b=1}^{2} 4\epsilon \left[ \left( \frac{\sigma}{r_{ab}} \right)^{12} - \left( \frac{\sigma}{r_{ab}} \right)^6 \right]. \nonumber
\end{equation}
The quadrupolar contribution $u_{\rm Q}$, is given by \cite{gray84}
\begin{equation}
u_{\rm Q}(\bm{r}_{ij},\bm{\omega}_i,\bm{\omega}_j,Q)=\frac{3}{4}\frac{Q^2}{\left|\bm{r}_{ij}\right|^5} f_{\rm Q}\left(\bm{\omega}_i,\bm{\omega}_j\right).
\label{uQ}
\end{equation}
Herein, ${\bm r}_{ij}$ is the centre-centre distance vector of two molecules $i$ and $j$, $r_{ab}$ is one of the four Lennard-Jones site-site distances; $a$ counts the two sites of molecule $i$, $b$ counts those of molecule $j$. The vectors ${\bm \omega}_i$ and ${\bm \omega}_j$ represent the orientations of the two molecules $i$ and $j$. $f_{\rm Q}$ is a trigonometric function depending on these molecular orientations, cf. \cite{gray84}. The Lennard-Jones parameters $\sigma$ and $\epsilon$ represent size and energy, respectively.
2CLJQ models have the four state independent model parameters $\sigma$, $\epsilon$, $L$, and $Q$, which have been adjusted to experimental vapour pressure, bubble density and critical temperature in a recent work of our group \cite{vraquad}. All pure substance parameters are given in Table 1. The spherical non-polar LJ model for methane is a limiting case of 2CLJQ models, where $L$ = 0 and $Q$ = 0.

To perform simulations of mixtures, a molecular mixture model is needed. On the basis of pairwise additive pure fluid models, molecular modelling of mixtures reduces to modelling interactions between unlike molecules. 
Here, the modified Lorentz-Berthelot combining rule with one adjustable binary interaction parameter $\xi$ was used for each unlike Lennard-Jones interaction 
\begin{equation}
\sigma_{ij}= {\frac{\sigma_{i} + \sigma_{j}}{2}}, \label{sig}
\end{equation}
\begin{equation}
\epsilon_{ij}= \xi \cdotp {\sqrt{\epsilon_{i} \cdotp \epsilon_{j}}}. \label{episi}
\end{equation}
The state independent parameter $\xi$ was adjusted to one experimental vapour pressure of each binary mixture in prior work of our group \cite{vlemix,modmix,huang}. 
Table 2 reports the six binary parameters of the quaternary natural gas mixture model.  
The unlike quadrupolar interactions are treated in a physically straightforward way, following the laws of electrostatics without any binary parameters.
It should be pointed out, that exclusively experimental VLE data were used in the parameterization of the molecular mixture model, but no caloric data. 

For the calculation of JT inversion curves on the basis of a given model by molecular simulation, several methods have been proposed in the literature
\cite{compress1,compress2,thermal,kiomag}. 
Here, as in prior work \cite{vrabecjt}, an intuitive and straightforward method was used. 
To calculate one JT inversion pressure $p$ for a given temperature $T$, a series of simulations, generally from 5 to 10, were made around the expected result, covering typically a rather large pressure range of 20 MPa. 
In these simulations, the enthalpy $h$ and its partial derivative with respect to pressure $(\partial h/\partial p)_{T,\bm{x}}$ were calculated. 
Both data sets were fitted simultaneously by a second order polynomial vs. pressure at that particular temperature. The inversion pressure corresponds simply to the minimum value of enthalpy, i.e. the minimum of that quadratic fit.

Molecular dynamics simulations were performed in the isobaric-isothermal ($N\!pT$) ensemble, using Andersen's barostat \cite{andersen1} and isokinetic velocity scaling \cite{allen1} for thermostating. 
After 6 000 equilibration time steps, the residual enthalpy \cite{heermann} 
\begin{eqnarray}
h^{res} = \frac{1}{N}.\left\lbrace <\sum_{i=1}^{N} \sum_{j>i}^{N}
u_{\rm 2CLJQ}(\bm{r}_{ij},\bm{\omega}_i,\bm{\omega}_j,L,Q)> + p<V>\right\rbrace  - k_{\rm B}T ,
\end{eqnarray} 
was averaged over 200 000 time steps, where the first term indicates the simulation average over the intermolecular potential energy and $<V>$ is the average of the extensive volume.
The partial derivative was obtained by a fluctuation expression \cite{hill} 
\begin{eqnarray}
\left( \frac{\partial h^{res}}{\partial p}\right) _{T,\bm{x}}= \frac{1}{N}.\left\lbrace
\frac{1}{k_{B}T}.\left[ < V >< H^{res} > - < VH^{res} > \right] + < V >\right\rbrace,
\end{eqnarray}
where $H^{res}$ is the extensive residual enthalpy. The ideal part of the enthalpy is not relevant, as it is not pressure dependent. 
 
Depending on the density of the state point, the membrane mass parameter $M$ of Andersen's barostat \cite{andersen1} was chosen from $10^{-20}$ to $10^{-15}$ kg/m$^4$. The intermolecular interactions were evaluated explicitly up to a cut-off radius of $r_c=5\sigma$ and standard long range corrections were used for the Lennard-Jones interaction, employing angle averaging as proposed by Lustig \cite{lustig}. 
Long-range corrections for the quadrupolar interaction are not needed since they disappear.
A total number of $N=1372$ molecules were initially placed in a fcc lattice configuration into a cubic simulation box, where the density of the system was chosen close to that expected from an EOS, if available, otherwise estimates were used. 

\section{Results}
JT inversion curves are compared for six different systematically chosen gas mixtures consisting of the four main natural gas components methane, nitrogen, carbon dioxide and ethane. 
Firstly, the three methane containing binary systems that can be formed from these four components were studied at equimolar composition, where the highest effect of mixing can be expected. 
Following this, secondly, all three ternary systems containing methane were studied, again at equimolar composition, i.e. with a mole fraction $x_i=1/3$ throughout. Finally, also the four pure fluids were investigated.

Simulation results for pure methane and carbon dioxide were taken from prior work \cite{vrabecjt}, but for nitrogen and ethane as well as for all mixtures new simulation results are presented. 
Tables 3 and 4 compile the full simulation data set.
The statistical uncertainties of this data are estimated to be in the order of 1 MPa. This translates into a relative error of around 2 \% for medium temperatures, but the relative errors are considerably larger at extremely low and high temperatures, where the inversion pressure approaches zero. 

For all systems the results for the JT inversion curves from molecular simulation are compared to the four mixture EOS, i.e. DDMIX \cite{ddmix1,ddmix2}, SUPERTRAPP \cite{supertrapp},  
BACKONE-EOS \cite{backone} and GERG-2004 \cite{newwagner}. 
The pure substance results are additionally compared to reference EOS using the program package REFPROP \cite{refprop}. 
All results are presented in pressure-temperature diagrams as well as in deviation plots.

\subsection{Pure components}
Figures 1 and 2 present the JT inversion curves of the pure fluids, which were grouped to achieve a good visibility. 
In these Figures, it can be seen that the results from the different methods qualitatively agree. 
Especially in the low temperature region of the JT inversion curves, they are often undistinguishable in these absolute plots.
Significant deviations, however, occur for higher temperatures. 
For a more detailed discussion, the deviation plots in Figure 3 are more suited, where the baselines represent GERG-2004. 
It should be pointed out that in Figure 3 additionally the pure substance reference EOS \cite{wagmeth,wagn2,wagco2,ethane_eos} results are shown. 
GERG-2004 agrees with the reference EOS within less than about 3 \% throughout, except for methane at high temperatures. 
DDMIX and BACKONE agree roughly equally well to GERG-2004 as well as the simulation data, often with the same trends. 
Generally, these results lie within a band of about 5 \%, larger deviations are found particularly at high temperatures.
BACKONE yields mostly higher results for high temperatures than GERG-2004, whereas DDMIX tends to yield lower values.
Significantly worse is SUPERTRAPP which yields negative deviations of more than 10 \% in large parts of the high temperature range of methane and nitrogen.

\subsection{Mixtures}

Figures 4 to 6 present the JT inversion curves for the six equimolar binary and ternary mixtures, which are again grouped to achieve good visibility. 
As for the pure fluids, the qualitative agreement between the different studied models is observed throughout. 
The agreement between all models is almost always excellent for low temperatures, but significant deviations are found in the high temperature region. 
Compared to the pure substances the spread between the different models is larger. 
Also here, the results can better be resolved in deviation plots, where the baselines were again chosen to represent GERG-2004 data. 
Figure 7 shows the binary and Figure 8 the ternary cases.  
Simulation data and GERG-2004 agree also for mixtures usually within 5 \%, larger relative deviations are found for very high temperatures. 
BACKONE shows a comparable performance. It yields systematically higher values at strongly elevated temperatures and a better agreement with GERG-2004 than DDMIX and SUPERTRAPP.
These two EOS show poorer results throughout, which are always too low by more than 10 \% at high temperatures. 
SUPERTRAPP yields the largest deviations in all cases.  

GERG-2004 was taken as a reference here as the broadest possible experimental data set was taken for its parameterization. 
This is confirmed by the fact that GERG-2004 data lies almost always in between the remaining results. 
Comparing it to the reference EOS for pure fluids, which were individually fitted including JT inversion data, an uncertainty of only about 3 \% has to be assumed, except for methane at high temperatures.

\section{Conclusion}
In this work, results from molecular modelling and simulation were compared to four advanced mixture EOS regarding JT inversion curves of natural gas mixtures. With a focus on the four most important components methane, nitrogen, carbon dioxide and ethane, six different equimolar binary and ternary mixtures were selected systematically.
As a reference and for completeness, the four components were also regarded as pure fluids. 
The comparison shows that molecular modelling and simulation is competitive with the most state-of-the-art EOS in predicting JT inversion curves. 
This approach has a similar performance as BACKONE, which is based on molecular simulation data itself.  
DDMIX and particularly SUPERTRAPP are less reliable. 

It can be stated that for other mixtures, where no such elaborate EOS are available, that molecular modelling and simulation is the method of choice to predict JT inversion. 
 
\section{Acknowledgment}
We gratefully acknowledge financial support by Deutsche Forschungsgemeinschaft, Schwerpunktprogramm 1155.

\clearpage

\noindent
{\bf List of symbols}

\begin{tabular}{ll}
$a$ & interaction site index \\
$b$ & interaction site index \\
$c_p$ & molar isobaric heat capacity \\ 
$f_Q$ & short notation for a trigonometric function \\
$h$ & molar enthalpy \\
$H$ & extensive enthalpy \\
$k_B$ & Boltzmann constant \\
$L$ & molecular elongation \\
$N$ & number of molecules \\
$p$ & pressure \\
$Q$ & molecular quadrupole momentum \\
$r_{ab}$ & site-site distance \\
$r_{\rm c}$ & center-center cut-off radius \\
$T$ & temperature \\
$u$ & pair potential \\
$V$ & extensive volume \\
$x$ & mole fraction \\
$\epsilon$ & Lennard-Jones energy parameter \\
$\mu$ & Joule-Thomson coefficient \\
$\xi$ & binary interaction parameter \\
$\sigma$  & Lennard-Jones size parameter \\
\end{tabular}

\noindent
\textbf{Vector properties} \\[0.1cm]
\begin{tabular}{ll}
$\bm{r}$ & distance vector \\
$\bm{x}$ & mole fraction vector \\
$\bm{\omega}$ & orientation vector \\
\end{tabular}

\clearpage

\noindent
\textbf{Subscript} \\[0.1cm]
\begin{tabular}{ll}
$a$ & interaction site index \\
$b$ & interaction site index \\
$i$ & molecule index \\
$i$ & component index \\
$j$ & molecule index \\
$j$ & component index \\
Q & point quadrupole \\
2CLJ & two-center Lennard-Jones \\
2CLJQ & two-center Lennard-Jones plus point quadrupole \\
\end{tabular}

\noindent
\textbf{Superscript} \\[0.1cm]
\begin{tabular}{ll}
$res$   & residual property \\
\end{tabular}

\clearpage

\begin{table}[ht]
\noindent
\caption[]{Parameters of pure fluid molecular models, taken from

\cite{vraquad}.}
\label{tab1}
\bigskip
\begin{center}
\begin{tabular}{lcccc} \hline
Fluid            & $\sigma/$\r{A}  &  $\left(\epsilon/k_{\rm B}\right)/$K  &  $L/$\r{A}  &  $Q/$ D\r{A} \\ \hline\hline
methane          & 3.7281 & 148.55  & -      & -      \\ 
nitrogen         & 3.3211 & 34.897  & 1.0464 & 1.4397 \\ 
carbon dioxide   & 2.9847 & 133.22  & 2.4176 & 3.7938 \\

ethane           & 3.4896 & 136.99  & 2.3762 & 0.8277 \\  \hline
\end{tabular}
\end{center}
\end{table}

\bigskip
\bigskip
\bigskip
\bigskip
\bigskip
\bigskip

\begin{table}[ht]
\noindent
\caption[]{Binary interaction parameters, taken from \cite{vlemix,modmix,huang}.}
\label{tab2}
\bigskip
\begin{center}
\begin{tabular}{lc}  \hline
Mixture        	          	           &   ${\xi}$	\\ \hline\hline
methane         +  carbon dioxide	   &	0.997	\\ 
methane         +  ethane       	   & 	0.958	\\ 
nitrogen        +  methane 	           & 	0.974	\\ 
nitrogen        +  carbon dioxide	   & 	0.962	\\ 

nitrogen        +  ethane	           & 	0.954	\\ 
carbon dioxide  +  ethane	           & 	1.041	\\ \hline
\end{tabular}
\end{center}
\end{table}
\clearpage

\begin{table}[ht]
\noindent
\caption[]{Molecular simulation results for JT inversion of the pure components. Data for methane and carbon dioxide was taken
from \cite{vrabecjt}, the remainder is from this work.}
\label{tab3}
\bigskip

\begin{center}
\begin{tabular}{cc|cc|cc|cc}   \hline
$T$ / K & $p$ / MPa & $T$ / K & $p$ / MPa & $T$ / K & $p$ / MPa & $T$ / K & $p$ / MPa  \\\hline \hline
\multicolumn{2}{l|}{methane} &	\multicolumn{2}{l|}{nitrogen} & \multicolumn{2}{l|}{carbon dioxide}	&\multicolumn{2}{l}{ethane} \\ 
178.26	&	15.83	&	100	&	\01.81	&	\0300	&	29.05	&	\0250	&	\02.83	\\
222.83	&	28.43	&	125	&	12.31	&	\0350	&	58.80	&	\0275	&	12.50	\\
267.39	&	41.22	&	150	&	24.68	&	\0400	&	76.09	&	\0300	&	21.58	\\
311.96	&	47.53	&	175	&	31.13	&	\0450	&	82.32	&	\0325	&	29.42	\\
356.52	&	50.83	&	200	&	35.78	&	\0500	&	87.16	&	\0350	&	34.94	\\
401.09	&	51.73	&	250	&	39.54	&	\0550	&	90.27	&	\0375	&	40.45	\\
445.65	&	51.44	&	300	&	38.88	&	\0600	&	91.31	&	\0425	&	48.80	\\

490.22	&	50.58	&	350	&	36.72	&	\0650	&	90.62	&	\0475	&	53.93	\\
534.78	&	48.73	&	400	&	31.55	&	\0700	&	87.16	&	\0500	&	57.38	\\
579.35	&	45.68	&	450	&	24.62	&	\0750	&	80.94	&	\0525	&	58.28	\\
623.91	&	42.81	&	500	&	17.37	&	\0800	&	76.09	&	\0600	&	60.14	\\
668.48	&	39.56	&	550	&	\08.48	&	\0850	&	69.18	&	\0675	&	60.82	\\
713.04	&	33.62	&		&		&	\0900	&	60.87	&	\0750	&	56.97	\\
757.61	&	27.73	&		&		&	\0950	&	53.96	&	\0825	&	54.34	\\
802.17	&	22.68	&		&		&	1000	&	45.66	&	\0900	&	48.37	\\
846.74	&	15.27	&		&		&	1050	&	37.35	&	\0975	&	42.13	\\
891.30	&	10.68	&		&		&	1100	&	28.36	&	1000	&	38.86	\\
935.87	&	\06.32	&		&		&	1150	&	17.99	&		&		\\

965.58	&	\03.17	&		&		&	1200	&	\08.30	&		&		\\ \hline
\end{tabular}
\end{center}
\end{table}
\clearpage

\begin{table}[ht]
\noindent
\caption[]{Molecular simulation results for JT inversion of the mixtures, this work.}
\label{tab4}
\bigskip
\begin{center}
\begin{tabular}{cc|cc|cc|cc}   \hline
$T$ / K & $p$ / MPa & $T$ / K & $p$ / MPa & $T$ / K & $p$ / MPa & $T$ / K & $p$ / MPa  \\\hline \hline
\multicolumn{2}{l|}{methane + carbon dioxide} &	\multicolumn{2}{l|}{methane + ethane}	& \multicolumn{2}{l|}{nitrogen + methane} &\multicolumn{2}{l}{methane +} \\
275	&	40.30	&	\0275	&	29.10	&	\0125	&	\00.80	& \multicolumn{2}{l}{carbon dioxide +} \\
350	&	58.68	&	\0350	&	45.14	&	\0200	&	31.28	& \multicolumn{2}{l}{nitrogen}	\\
425	&	68.17	&	\0425	&	53.33	&	\0275	&	43.44	&	200	&	22.87	\\
500	&	70.13	&	\0500	&	58.24	&	\0350	&	46.36	&	250	&	39.76	\\
575	&	66.90	&	\0575	&	58.26	&	\0425	&	42.24	&	300	&	51.44	\\
650	&	63.22	&	\0625	&	57.86	&	\0500	&	37.35	&	350	&	57.41	\\
725	&	55.81	&	\0700	&	55.53	&	\0575	&	28.46	&	425	&	59.35	\\
800	&	45.20	&	\0775	&	48.55	&	\0650	&	17.46	&	500	&	58.68	\\ \cline{5-6}
875	&	34.86	&	\0850	&	40.73	&	\multicolumn{2}{l|}{methane +}&	575	&	51.29	\\
950	&	23.07	&	\0925	&	33.50	&	\multicolumn{2}{l|}{carbon dioxide +}	&	650	&	43.62	\\ \cline{1-2}
\multicolumn{2}{l|}{methane +}&	\0950	&	29.86	&	\multicolumn{2}{l|}{ethane +}           &	725	&	30.74	\\
\multicolumn{2}{l|}{ethane +}&	1050	&	16.44	&       \0275	&	31.62	&	800	&	21.71	\\
\multicolumn{2}{l|}{nitrogen +}&	&		&	\0350	&	49.72	&	875	&	15.03	\\ 
200	&	17.46	&		&		&	\0425	&	60.60	&		&		\\
275	&	38.53	&		&		&	\0500	&	65.13	&		&		\\
350	&	48.64	&		&		&	\0575	&	65.15	&		&		\\
425	&	52.45	&		&		&	\0625	&	62.31	&		&		\\
500	&	51.80	&		&		&	\0700	&	59.16	&		&		\\
575	&	48.77	&		&		&	\0775	&	52.53	&		&		\\
650	&	42.27	&		&		&	\0850	&	45.07	&		&		\\
725	&	35.46	&		&		&	\0925	&	35.76	&		&		\\
800	&	25.26	&		&		&	1000	&	25.69	&		&		\\ \hline
\end{tabular}
\end{center}
\end{table}
\clearpage

\listoffigures
\clearpage

\begin{figure}[ht]
\begin{center}
\includegraphics[width=\textwidth]{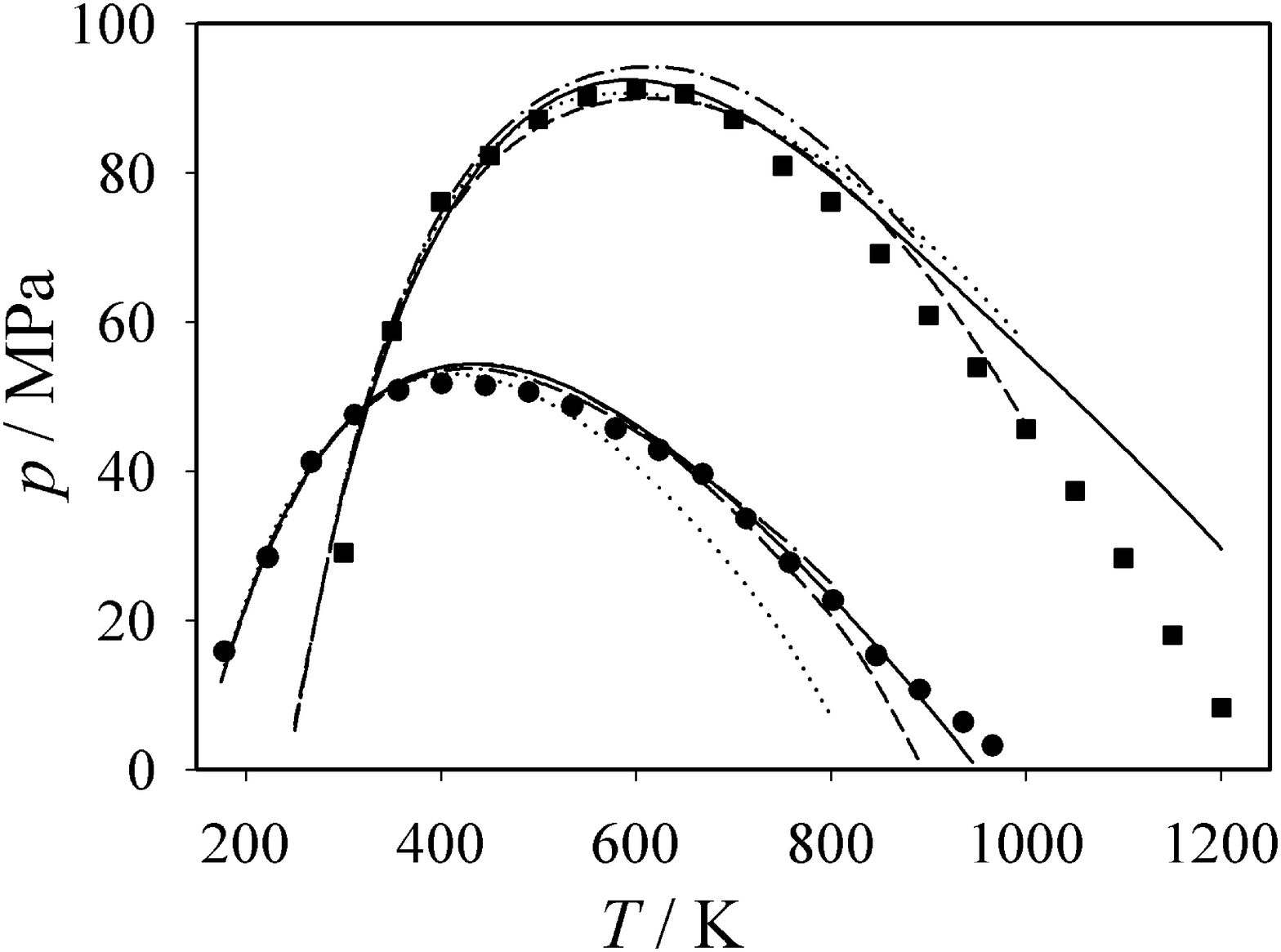}\
\end{center}
\caption[Joule-Thomson inversion curves of two pure fluids. Simulation: {\large $\bullet$} methane, taken from \cite{vrabecjt}, {\footnotesize $\blacksquare$} carbon
dioxide, taken from \cite{vrabecjt}; EOS: -~-~- DDMIX
\cite{ddmix1,ddmix2}, ... SUPERTRAPP \cite{supertrapp}, -.-.- BACKONE \cite{backone}, --- GERG-2004 \cite{newwagner}.]{Vrabec et al.} 
\label{f1}
\end{figure}
\clearpage

\begin{figure}[ht]
\begin{center}
\includegraphics[width=\textwidth]{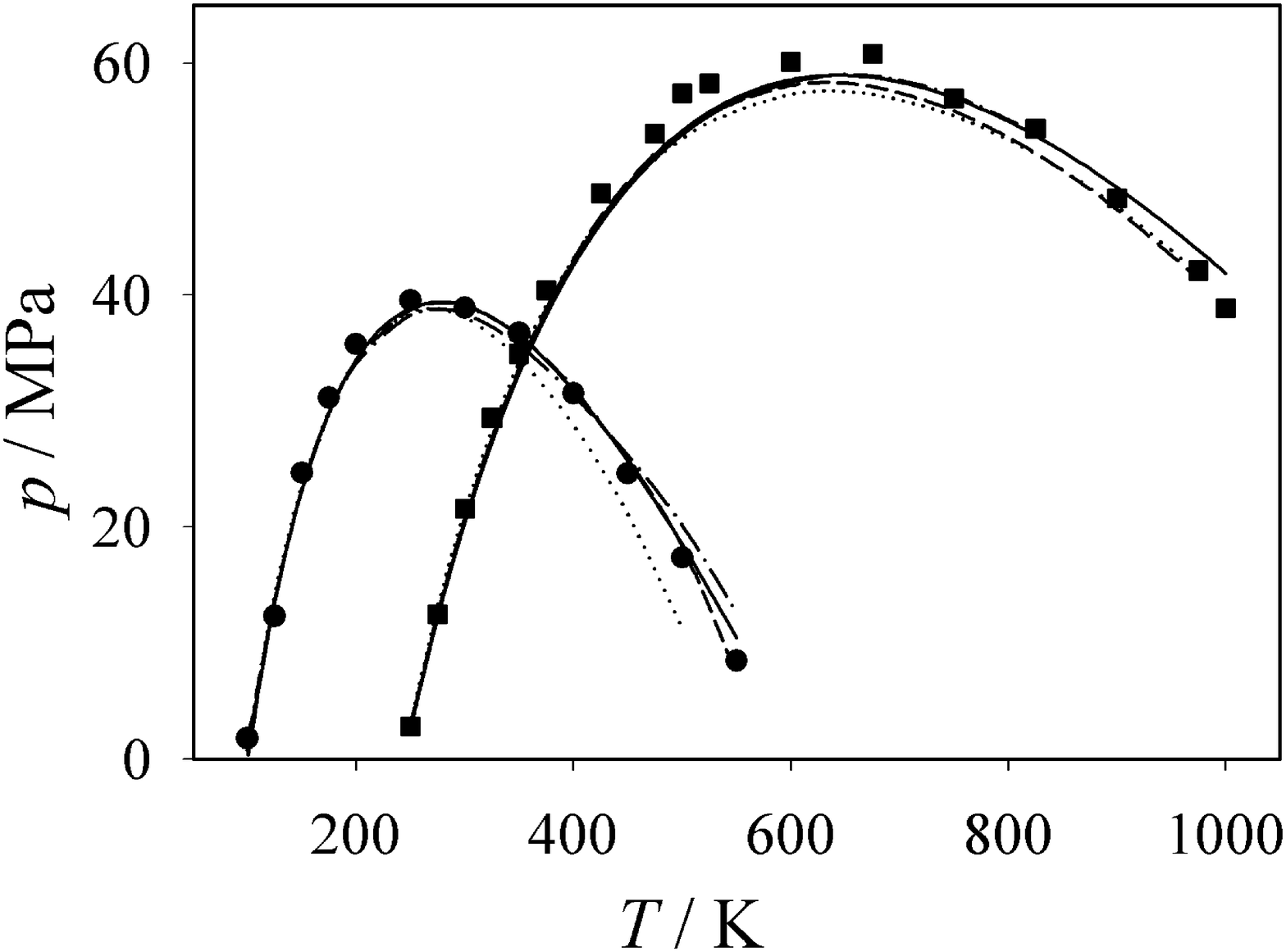}\
\end{center}
\caption[Joule-Thomson inversion curves of two pure fluids. Simulation: {\large $\bullet$} nitrogen, this work, {\footnotesize $\blacksquare$} ethane, this work; EOS:
-~-~- DDMIX \cite{ddmix1,ddmix2}, ... SUPERTRAPP \cite{supertrapp}, -.-.- BACKONE \cite{backone}, --- GERG-2004 \cite{newwagner}.]{Vrabec et al.} 
\label{f2}
\end{figure}
\clearpage

\begin{figure}[ht]
\begin{center}
\includegraphics[width=\textwidth]{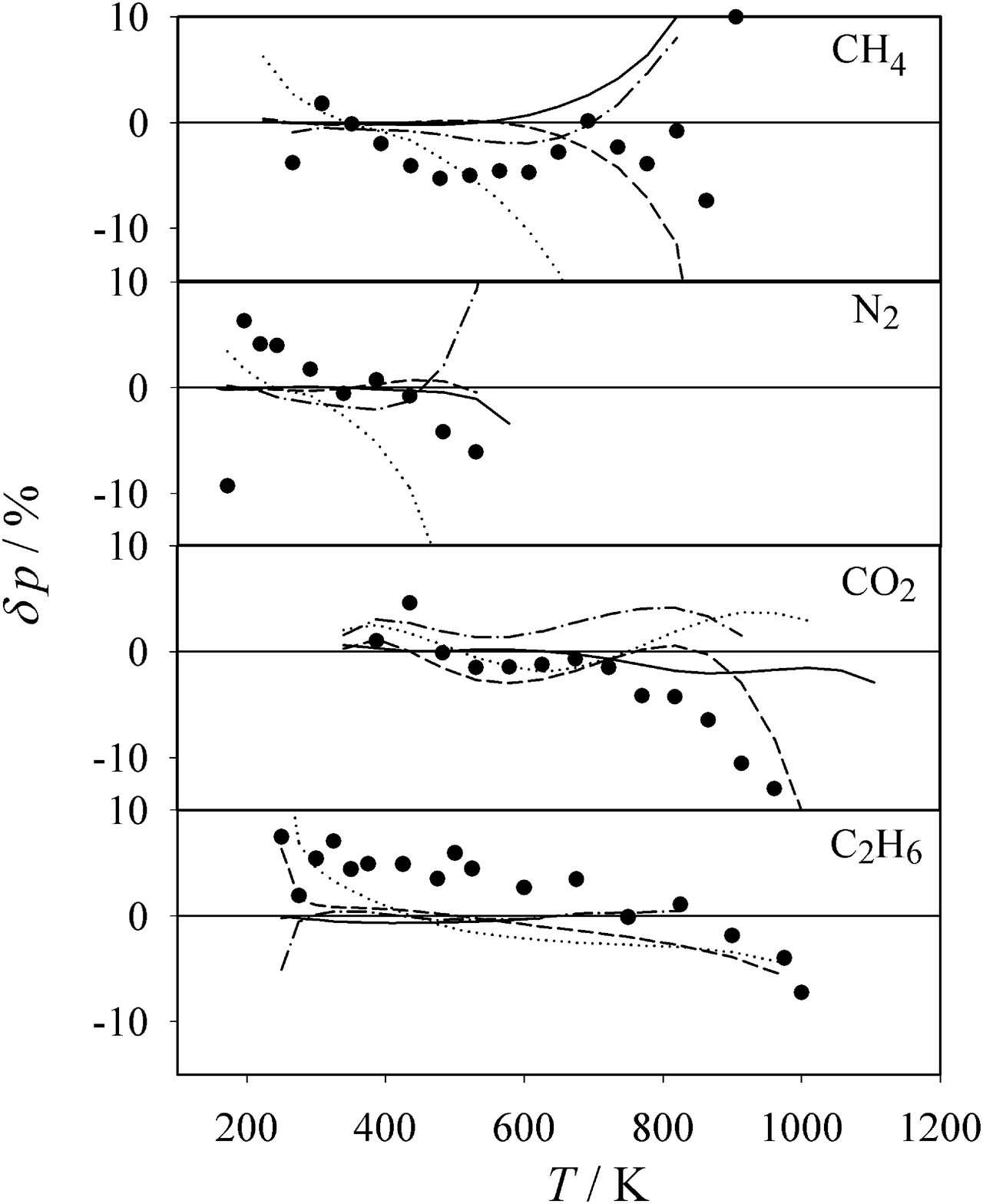}\
\end{center}
\caption[Deviation plots for Joule-Thomson inversion of the pure fluids. The baselines represent GERG-2004 \cite{newwagner}. Simulation: {\large $\bullet$}; EOS: -~-~- DDMIX \cite{ddmix1,ddmix2}, ...
SUPERTRAPP \cite{supertrapp}, -.-.- BACKONE \cite{backone}; Reference EOS:
 --- Setzmann and Wagner \cite{wagmeth} (methane), Span et al. \cite{wagn2} (nitrogen), Span and Wagner \cite{wagco2} (carbon dioxide)
and Friend et al. \cite{ethane_eos} (ethane).]{Vrabec et al.} 
\label{f3}
\end{figure}
\clearpage

\begin{figure}[ht]
\begin{center}
\includegraphics[width=\textwidth]{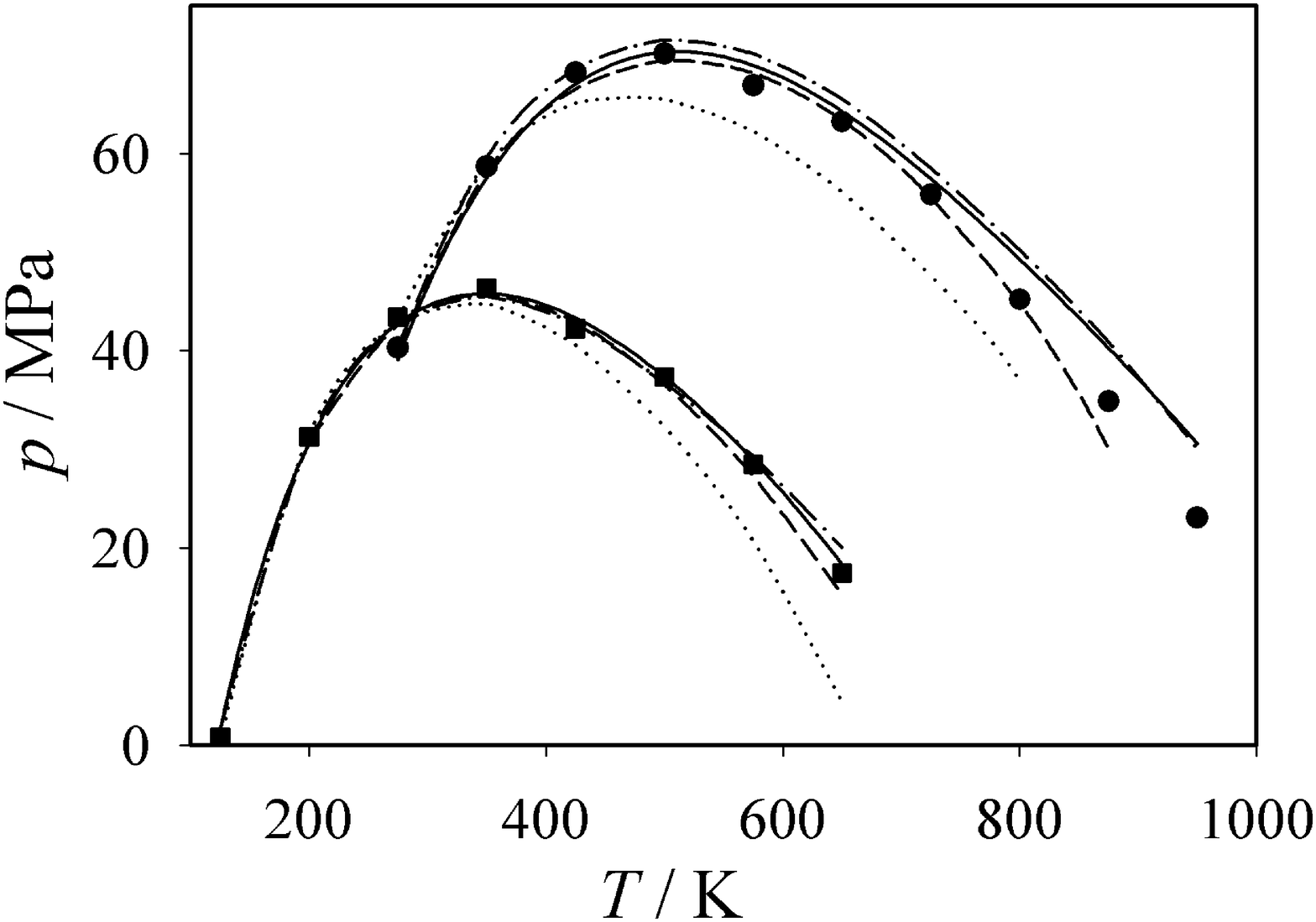}\
\end{center}
\caption[Joule-Thomson inversion curves of two equimolar mixtures. Simulation: {\large $\bullet$} methane + carbon dioxide, this work, {\footnotesize $\blacksquare$} nitrogen +
methane, this work; EOS: -~-~- DDMIX \cite{ddmix1,ddmix2}, ... SUPERTRAPP \cite{supertrapp}, -.-.- BACKONE \cite{backone}, --- GERG-2004 \cite{newwagner}.]{Vrabec et al.}
\label{f4}
\end{figure}
\clearpage

\begin{figure}[ht]
\begin{center}
\includegraphics[width=\textwidth]{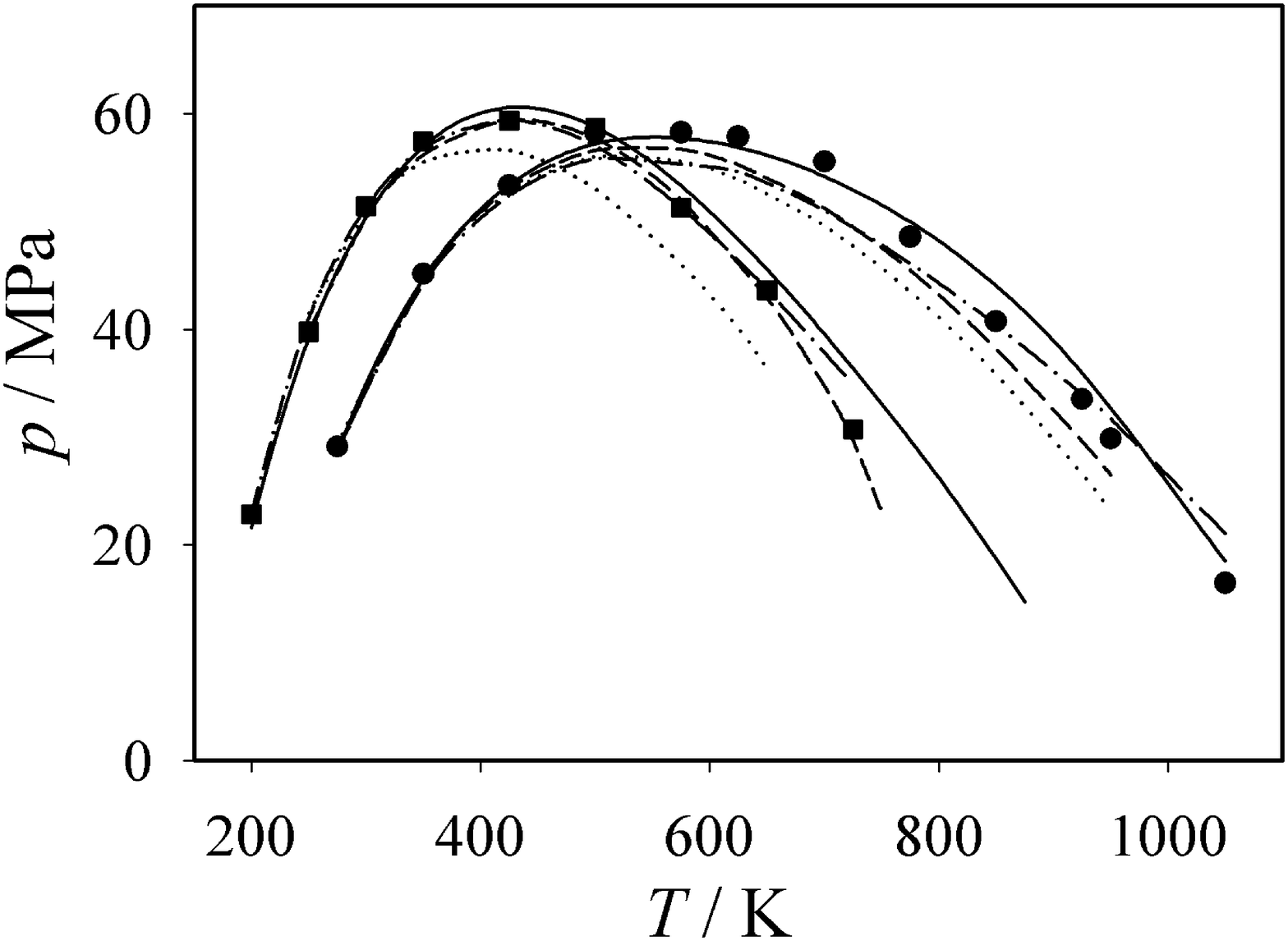}\
\end{center}
\caption[Joule-Thomson inversion curve of two equimolar mixtures. Simulation: {\large $\bullet$} methane + ethane, this work, {\footnotesize $\blacksquare$} methane + carbon dioxide + nitrogen, this work; EOS: -~-~- DDMIX \cite{ddmix1,ddmix2}, ... SUPERTRAPP \cite{supertrapp}, -.-.- BACKONE \cite{backone}, --- GERG-2004 \cite{newwagner}.]{Vrabec et al.}
\label{f5}
\end{figure}
\clearpage

\begin{figure}[ht]
\begin{center}
\includegraphics[width=\textwidth]{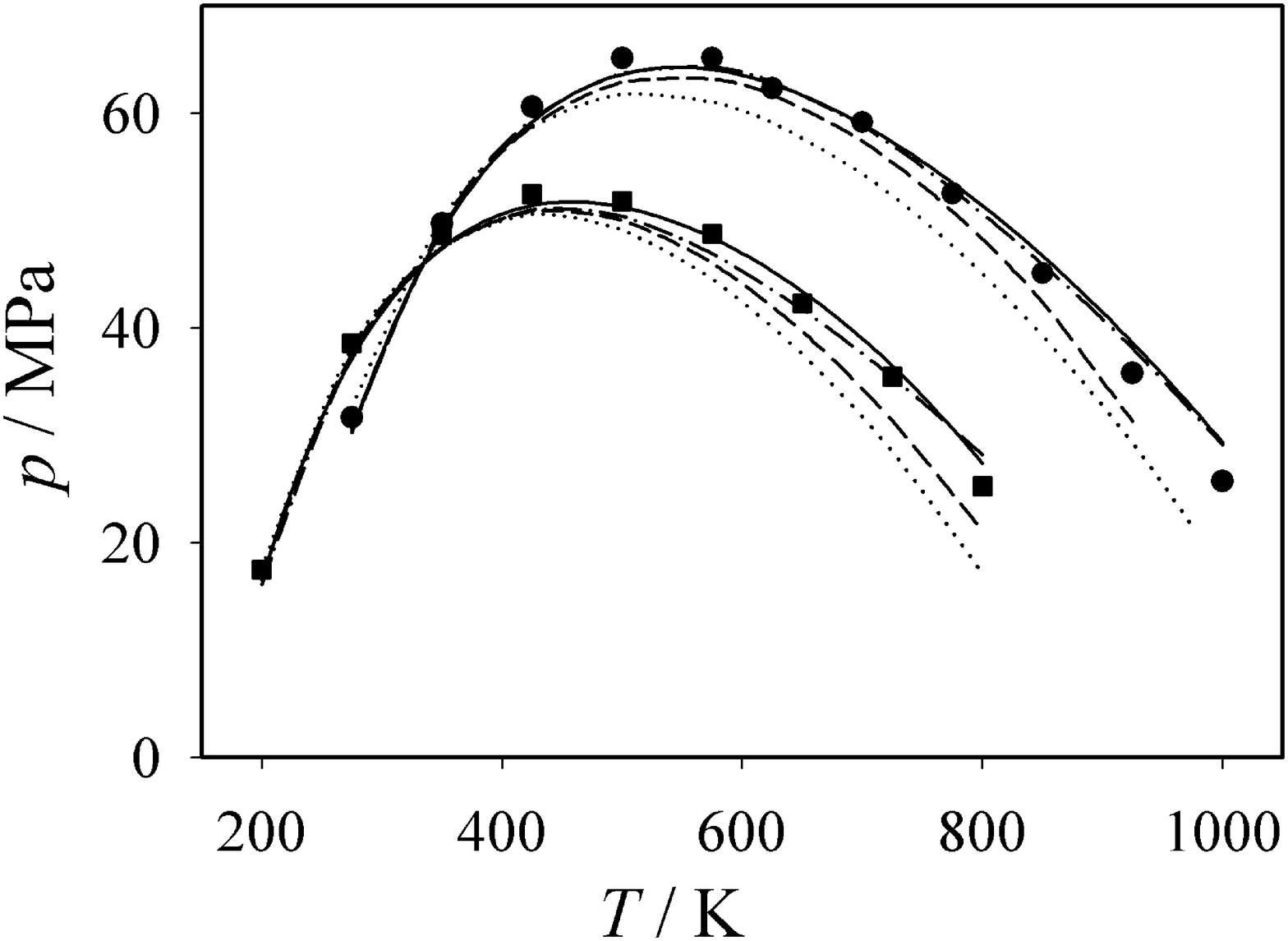}\
\end{center}
\caption[Joule-Thomson inversion curve of two equimolar mixtures. Simulation: {\large $\bullet$} methane + carbon dioxide + ethane, this work, {\footnotesize $\blacksquare$} methane + ethane + nitrogen, this work; EOS: -~-~- DDMIX \cite{ddmix1,ddmix2}, ... SUPERTRAPP \cite{supertrapp}, -.-.- BACKONE \cite{backone}, --- GERG-2004 \cite{newwagner}.]{Vrabec et al.}
\label{f6}
\end{figure}
\clearpage

\begin{figure}[ht]
\begin{center}
\includegraphics[width=\textwidth]{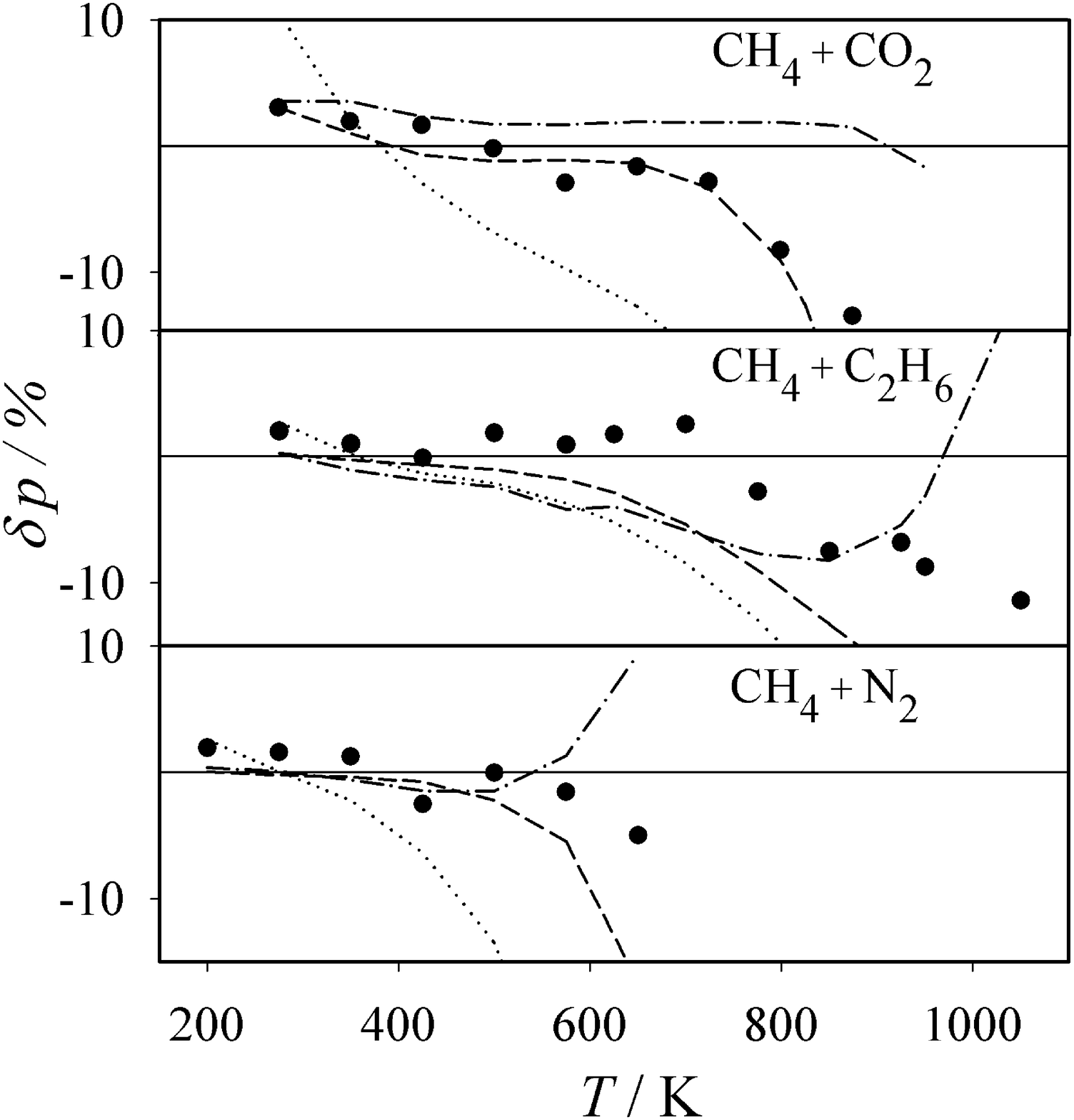}\
\end{center}
\caption[Deviation plots for Joule-Thomson inversion of the binary mixtures. The baselines represent GERG-2004 \cite{newwagner}. Simulation: {\large $\bullet$}; EOS: -~-~- DDMIX \cite{ddmix1,ddmix2}, ... SUPERTRAPP \cite{supertrapp}, -.-.- BACKONE \cite{backone}.]{Vrabec et al.}
\label{f7}
\end{figure}
\clearpage

\begin{figure}[ht]
\begin{center}
\includegraphics[width=\textwidth]{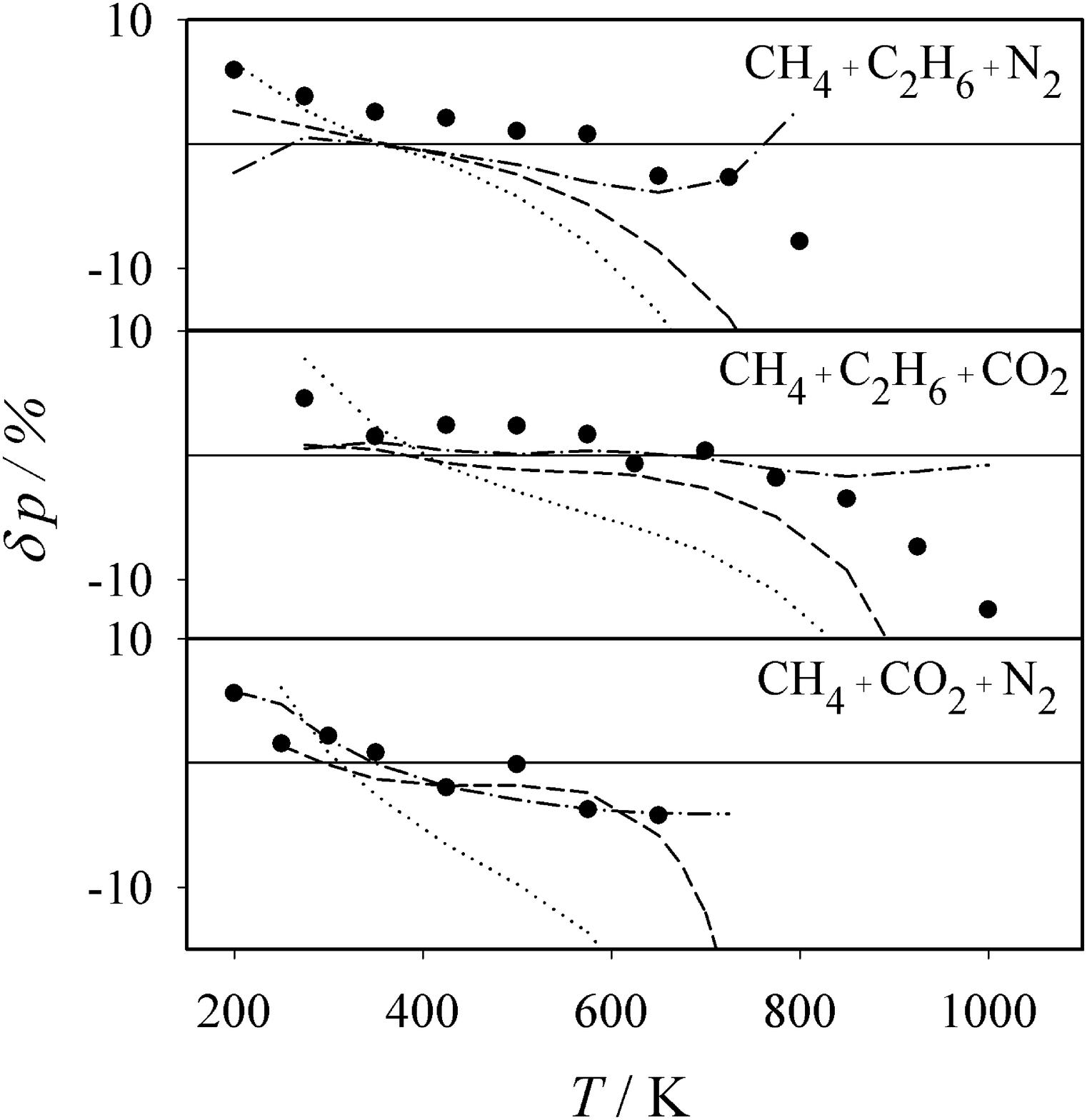}\
\end{center}
\caption[Deviation plots for Joule-Thomson inversion of the ternary mixtures. The baselines represent GERG-2004 \cite{newwagner}. Simulation: {\large $\bullet$}; EOS: -~-~- DDMIX \cite{ddmix1,ddmix2}, ... SUPERTRAPP \cite{supertrapp}, -.-.- BACKONE \cite{backone}.]{Vrabec et al.}
\label{f8}
\end{figure}
\clearpage

\end{document}